\documentclass[proof]{pasj01}
\Received{$\langle$reception date$\rangle$}
\Accepted{$\langle$acception date$\rangle$}
\Published{$\langle$publication date$\rangle$}

\usepackage{color}

\usepackage{ulem}

\usepackage[outdir=./]{epstopdf}

\newcommand{\pow}[2]{\ensuremath{{#1}\times 10^{#2}}}
\newcommand{\nucm}[2]{\ensuremath{{}^{#1}{\rm #2}}}

\newcommand{\feoh}{\textrm{[Fe/H]}}
\newcommand{\ali}{\ensuremath{A(\textrm{Li})}}
\newcommand{\msun}{\ensuremath{M_{\odot}}}
\newcommand{\rsun}{\ensuremath{R_{\odot}}}
\newcommand{\teff}{\ensuremath{T_{\textrm {eff}}}}
\newcommand{\logg}{\ensuremath{\log {\textrm g}}}
\newcommand{\hyp}{\ensuremath{\, \mathchar`- \,}}

\begin{document}

\title{First Star Survivors as Metal-Rich Halo Stars that Experienced Supernova Explosions in Binary Systems}
\author{Takuma Suda$^{1,2,3}$}%
\author{Takayuki R. Saitoh$^{4,5}$}%
\author{Yuki Moritani$^{6}$}%
\author{Tadafumi Matsuno$^{7}$}%
\author{Toshikazu Shigeyama$^{2}$}%
\altaffiltext{1}{Department of Liberal Arts, Tokyo University of Technology,
 5-23-22 Kamata, Ota-ku, Tokyo 144-8535, Japan }
 \altaffiltext{2}{Research Center for the Early Universe, The University of Tokyo,
 7-3-1 Hongo, Bunkyo-ku, Tokyo 113-0033, Japan }
\altaffiltext{3}{The Open University of Japan,
 2-11 Wakaba, Mihama-ku, Chiba, Chiba 261-8586, Japan }
\altaffiltext{4}{Department of Planetology, Graduate School of Science, Kobe University,
 1-1 Rokkodai-cho, Nada-ku, Kobe, Hyogo 657-8501, Japan }
\altaffiltext{5}{ Earth-Life Science Institute, Tokyo Institute of Technology,
 2-12-1 Ookayama, Meguro-ku, Tokyo 152-8551, Japan }
 \altaffiltext{6}{Kavli Institute for the Physics and Mathematics of the Universe (WPI), The University of
Tokyo,
 5-1-5 Kashiwanoha, Kashiwa, Chiba 277-8583, Japan }
 \altaffiltext{7}{Kapteyn Astronomical Institute, University of Groningen,
 Landleven 12, 9747 AD Groningen, The Netherlands }
 \email{sudatkm@stf.teu.ac.jp}


\maketitle

\begin{abstract}
The search for the first stars formed from metal-free gas in the universe is one of the key issues in astronomy because it relates to many fields, such as the formation of stars and galaxies, the evolution of the universe, and the origin of elements.
It is not still clear if metal-free first stars can be found in the present universe.
These first stars are thought to exist among extremely metal-poor stars in the halo of our Galaxy.
Here we propose a new scenario for the formation of low-mass first stars that have survived until today and observational counterparts in our Galaxy.
The first stars in binary systems, consisting of massive- and low-mass stars, are examined using stellar evolution models, simulations of supernova ejecta colliding with low-mass companions, and comparisons with observed data.
These first star survivors will be observed as metal-rich halo stars in our Galaxy.
We may have identified a candidate star in the observational database where elemental abundances and kinematic data are available.
Our models also account for the existence of several solar-metallicity stars in the literature having space velocities equivalent to the halo population.
The proposed scenario demands a new channel of star formation in the early universe and is a supplementary scenario for the origin of the known metal-poor stars.
\end{abstract}

\section{Introduction}
The first stars in the universe draw much attention for the understanding of the star formation history in the early Galaxy.
It involves the critical questions; do they still exist? how can they be found?
Answers to these questions have a strong impact on how our Sun was formed and where the terrestrial elements originated.
This is because the first stars give a hint to understand the formation and evolution of stars, and the synthesis of elements in stars, starting from elemental abundances equivalent to those in the big bang nucleosynthesis.

The existence of the first stars in the current universe is still controversial.
The basic understanding is that they are typically too massive to exist for a period of more than 10 Myr from their birth.
The preference for the formation of massive stars is due to the lack of elements that work as a coolant to compress star-forming gas \citep{Bromm2004}.
However, there are some arguments about the initial mass of the first stars.
There is a possibility of the formation of a star with the initial mass below a solar mass that has a lifetime comparable to or larger than the age of the universe \citep{Clark2011,Susa2014,Stacy2016}.
In particular, recent simulation studies of the first stars focus on the formation of such low-mass first stars around the massive stars \citep{Susa2014} or the formation of the first binary systems \citep{Stacy2016,Sugimura2020}.

The effort to find the first stars in our Galaxy has revealed hundreds of stars with $\feoh \lesssim -3$ \citep{Bond1980,Beers1985,Ryan1991,McWilliam1995b,Ryan1996,Fulbright2000,Norris2001,Carretta2002,Johnson2002,Cayrel2004,Cohen2004,Honda2004b,Barklem2005,Aoki2007b,Lai2008,Caffau2013,Norris2013,Roederer2014a,Francois2018}.
These stars are called extremely metal-poor (EMP) stars among which the currently most iron-poor star has $\feoh < -7$ \citep{Keller2014}.
The origin of EMP stars has been a controversial mainly in explaining the large fraction of carbon-enhanced metal-poor (CEMP) stars.
Previous scenarios can be classified into three theoretical models, all of which try to explain the abundance patterns of carbon-enhanced metal-poor (CEMP) stars.
In the three major scenarios, CEMP stars are formed (1) by binary mass transfer from the former AGB stars \citep{Suda2004}, (2) from the gas in the interstellar medium (ISM), contaminated by the ejecta from the first-generation supernova \citep{Umeda2003}, and (3) from the gas ejected from fast-rotating massive stars \citep{Meynet2006}.  

In this paper, we explore another possibility of the formation of the first stars that can be verified by observations, i.e., the evolution of the first stars in a binary system consisting of a massive star and a low-mass star.
In such a system, the ejecta of a supernova explosion in the massive star collide with its low-mass companion, which will either strip away or will be gravitationally confined to the surface of the companion or both may happen due to a wide range of the speed of the ejecta.
This scenario provides a new pathway to look for evidence of polluted first stars among known halo stars in the Milky Way galaxy, where there are more than 500 stars with detailed chemical abundances derived from high-resolution spectra.

The paper is organized as follows.
In the next section, we provide the overview of our scenario.
The details of the models are described in \S~\ref{sec:model}.
The results of our simulations are discussed in \S~\ref{sec:results}.
\S~\ref{sec:discussion} provides the implications from our scenario.
Conclusions follow in \S~\ref{sec:conclusion}.

\section{New scenario to identify the survivors of the first stars}

We propose that some low-mass first stars should have survived supernovae in binary systems.
They can be observed in our Galaxy as metal-rich halo stars if they were contaminated by the supernova ejecta in close binary systems.
A sufficiently small binary separation is required for surviving stars to change their surface chemical composition by the accretion of the ejecta.
This is not necessarily guaranteed because typical massive stars evolve to red supergiants, having radii up to $\sim 1000 \rsun$ before the core-collapse.
If the separation between two stars is shorter than the radius of the evolved massive star, the binary system will undergo the common envelope phase where the companion star will be embedded in the envelope of the massive star.
Such a system will have a short life span due to the mass transfer and cannot be a candidate for the first star binaries.

The first stars can avoid the evolution to red supergiants due to the initial lack of metals in the centre, more specifically CNO elements.
The final radius of a metal-free massive star is reported to be of the order of $10 \rsun$ \citep{Heger2010,Tanikawa2020}.
This result is supported even if stellar rotation is taken into account (K. Takahashi, T. Yoshida, priv. comm.).
The upper boundary of the metal content of stars not to evolve into red supergiants depends on the initial mass, where the threshold value is $\feoh \sim -7$ for $12 \msun$ and $\sim -2$ for $20 \msun$ (See \S~\ref{sec:evostar}).
It is to be noted that less massive stars are more abundant than more massive stars, and hence metal-free stars are dominant among massive stars with small final radii.

\begin{figure*}
 \begin{center}
  \includegraphics[width=160mm]{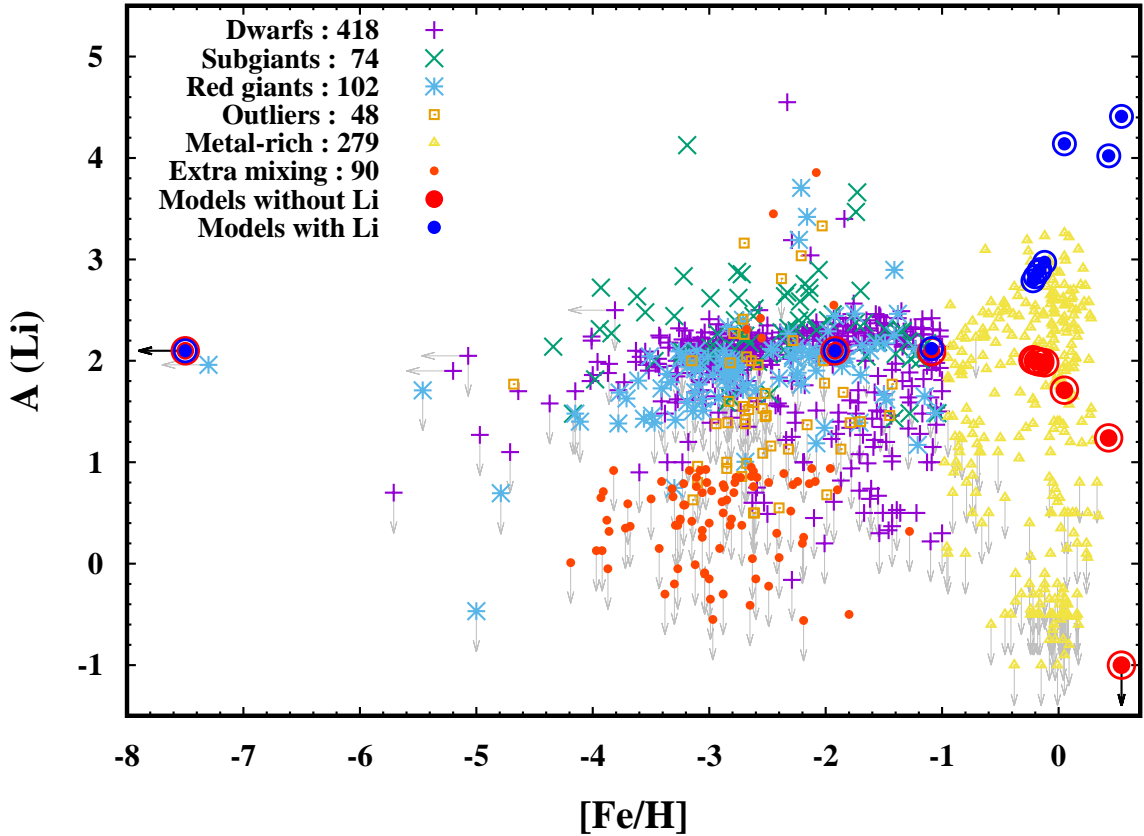}
 \end{center}
 \caption{Lithium abundances, $\ali = \log_{10} (N({\rm Li}) / N({\rm H}) + 12)$, as a function of metallicity, denoted by \feoh, where [A/B] $= \log_{10}( N_{\rm A} / N_{\rm B} )_{\rm star} - \log_{10}( N_{\rm A} / N_{\rm B} )_{\odot}$.
 The data points are compiled from the abundance data from selected literature using the SAGA database (see appendix).
 However, we made corrections to the abundances by considering the effect of dilution by the convective envelopes in subgiants and red giants to compare models and observations with lithium abundances at the main sequence phase.
 The plotted data are classified and counted (shown by the numbers next to the labels) according to the evolutionary status as discussed in the text.
 The data points with circles denote the model results in table~\ref{tab:model}.}\label{fig:life}
\end{figure*}

The evidence of the first supernova binaries can be checked observationally by lithium and iron abundances of companion stars.
Figure~\ref{fig:life} shows the lithium abundances of known metal-deficient stars as a function of metallicity.
Lithium is a good tracer of the change of the surface chemical composition because it is easily destroyed at layers with temperatures above $2.5 \times 10^{6}$ K, which means that the total lithium content is small.
Also, the depletion or enhancement of lithium is easily identified by observations because the stars in the main sequence phase have the typical value of lithium abundances of $\ali = 2.1$, which is called the Spite plateau \citep{Spite1982}.
Once the surface chemical composition is altered by external pollution such as stripping and/or accretion due to the collision of supernova ejecta, the surface lithium abundance is depleted by factors of several or more from the original value.
Observationally, a non-negligible fraction of metal-poor stars shows lithium depletion, which is not explained by the standard stellar evolution models.
In this study, we explore the possibility of lithium depletion.

Simulations of the collisions between supernova ejecta and a binary companion are performed using a three-dimensional Smoothed Particle Hydrodynamics (SPH) code, ASURA \citep{Saitoh2008}.
The initial conditions for the simulations are calculated by a one-dimensional Lagrangian hydrodynamics code calibrated by SN1987A \citep{Shigeyama1990} so that the model reproduces well the light curve of the supernova.
The initial mass and explosion energy are set at 15, 20, and $25 \msun$ and $10^{51}$ erg, respectively.
The remnant mass of the progenitor is assumed to be $1.3 \msun$, following the results of \citet{Heger2010}.
Stellar mass loss before the explosion is ignored, following the previous studies \citep{Heger2010,Tanikawa2020}.
The initial condition for the explosion is mapped on 3D simulations by setting the pressure, density, and kinetic energy in the primary star.
The structure of the secondary star is constructed by mapping a one-dimensional model star computed with a stellar evolution code \citep{Suda2010}.
We assumed that the convective zone of the surface of the model mixes the materials homogeneously and does not change its mass and structure by the collision of supernova ejecta.
This assumption may affect the estimate of the final chemical composition by the simulations.
Stellar rotation is ignored for the sake of simplicity.
In our simulation setups, binary components should be tidally synchronized so that the effect of rotation is not significant.

Only a separation is a relevant binary parameter in our simulations.
This is because the orbital velocity is much smaller than the impact velocity of the ejecta and because the simulation time is much shorter than the orbital period.
Here we presume that the eccentricity is close to zero to avoid the passage of a companion star into the envelope of a primary.
We fixed the supernova model in this study for the following reasons:
(1) the simulations by changing the velocity distribution and/or the total kinetic energy of the ejecta will be essentially the same as the models with different binary separations,
(2) it will be a big computation cost to run more simulations with different inputs for supernova models, and
(3) this project is still in embryo and it will be too much work to test unusual types of supernova such as asymmetric explosions and mixing and fallback models.
The impact of the variations in supernova models is addressed later in this paper and such simulations can be a future work.

In the next section, we describe the details of the models and assumptions to perform numerical simulations and compare the results with observations.

\section{Models and assumptions}\label{sec:model}

\subsection{Evolution of massive metal-free and metal-poor stars}\label{sec:evostar}

The characteristics of the evolution of metal-free and metal-poor stars are the smaller radii at the ends of their lives compared with the stars that evolve through the red supergiant phase. Metal-deficient massive stars are supported by hydrogen-burning in the centre at the early phase without CNO elements (the p-p chain reactions), which requires higher temperatures than hydrogen burning with CNO catalysts (the CNO cycles). The higher the temperature of the nuclear burning regions, the faster the progress of the nuclear burning stages. Therefore, the final nuclear burning finishes before the stars pass through the Hertzsprung gap where the radii increase rapidly.

\begin{figure}
 \begin{center}
  \includegraphics[width=80mm]{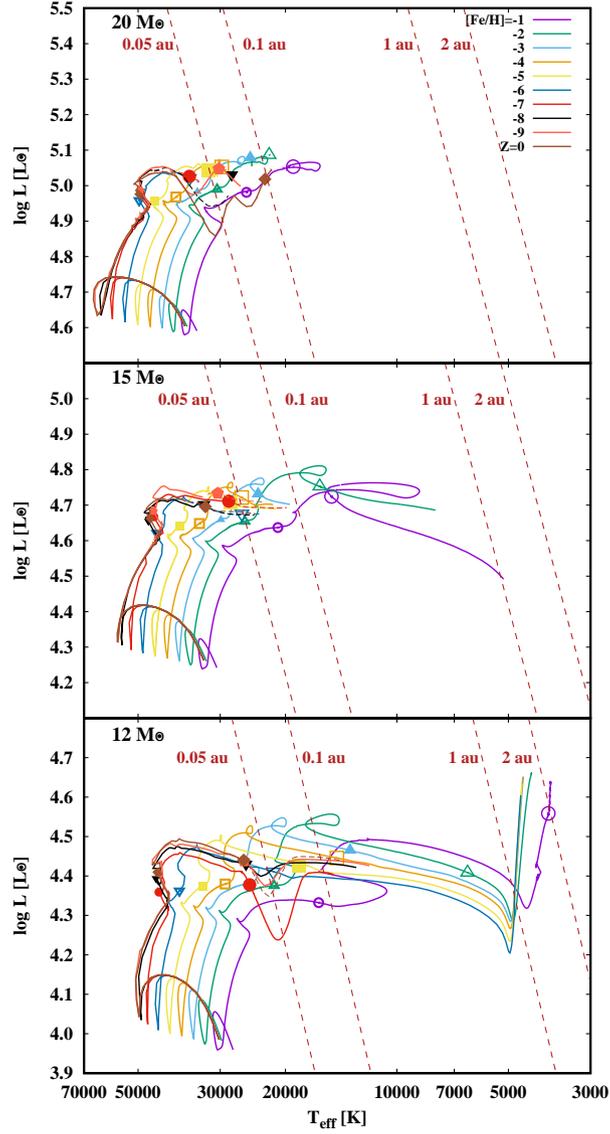}
 \end{center}
 \caption{Evolution of massive stars with $20 \msun$ (top panel), $15 \msun$ (middle panel), and $12 \msun$ (bottom panel) for various metallicities from metal-free ($Z = 0$) to $\feoh = -1$. The dashed lines represent the constant stellar radii on the H-R diagram. The two symbols on the evolutionary tracks correspond to the onset of helium (smaller symbols) and carbon (larger symbols) burning, respectively. Computations were terminated at the end of the carbon-burning phase where the mass fraction of carbon abundance is below 0.02.}\label{fig:hrd}
\end{figure}

We have computed the evolution of massive stars with various mass and metallicity, which is displayed in figure~\ref{fig:hrd}. We use the same stellar evolution code \citep{Suda2010} with the addition of carbon burning reactions. Cross-sections of $\nucm{12}{C} + \nucm{12}{C}$ and $\nucm{12}{C} + \nucm{16}{O}$ are taken from the fitting formulae provided in \citet{Caughlan1988}. The computations are terminated when the mass fraction of carbon in the centre becomes smaller than 0.02. The locations of the onset of the helium- and carbon-burning phases on the Hertzsprung-Russell (H-R) diagram are shown by smaller and larger circles, respectively. For the sake of simplicity, carbon and oxygen in these reactions are converted to \nucm{25}{Mg}. This approximation is sufficient to see the final location of model stars on the H-R diagram.

In some models with low-metallicity stars, we found numerical instabilities caused by the hydrogen ingestion into helium-burning layers where the temperature is the order of $10^{8}$ K.
For instance, $15 \msun$ models with the metallicity of $\feoh \leq -8$ experience the inward extension of hydrogen-burning convection into the helium core before the onset of carbon burning.
This results in the hydrogen flash with the hydrogen-burning luminosity exceeding $\log (L/L_{\odot}) > 10$, which does not provide any convergence in the stellar evolution.
It is not clear why this mixing occurs and should be studied in a separate paper.
The same phenomenon is reported in previous studies for rotating models with $M = 160 \msun$ \citep{Takahashi2018} and $M > 200 \msun$ \citep{Yoon2012}.
We tested models with different time steps and rezoning during these evolutionary phases, and we always encountered the same instabilities.
We also tested other models using different stellar evolution codes and found the same phenomena.

To compromise the computations of these models, we artificially prohibited the mixing of the convection into the helium core. This can be justified for the evolution of the star itself because the overall evolution is controlled by the nuclear burning at the centre, not by the shell burning. The evolutionary tracks in figure~\ref{fig:hrd} are apparently consistent with the previous studies \citep{Heger2010,Tanikawa2020}.

\begin{figure}
 \begin{center}
  \includegraphics[width=80mm]{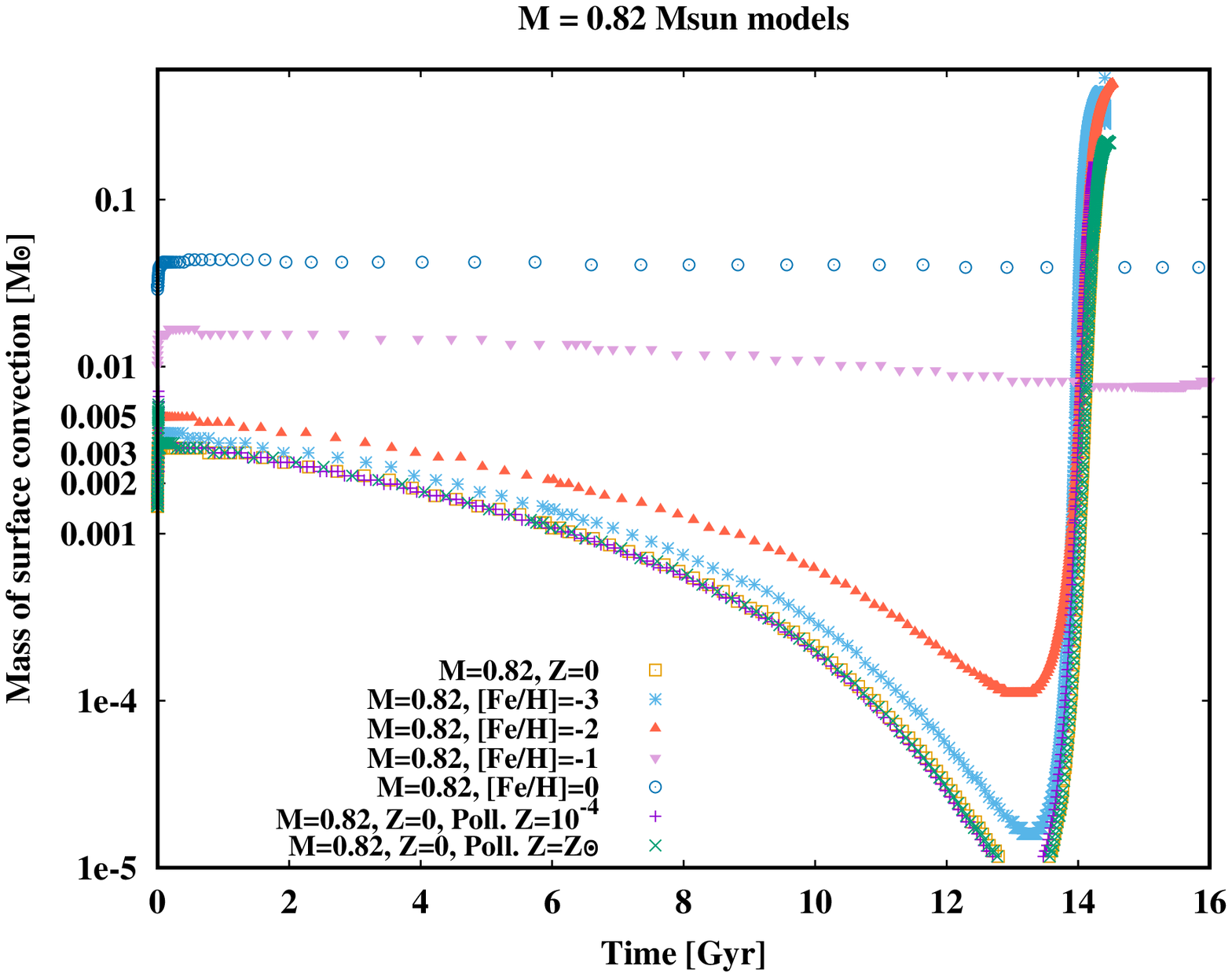}
 \end{center}
 \caption{Depth of the surface convective zones as a function of time for low-mass stars of $0.82 \msun$ with various metallicities. Surface pollution is also considered for two metal-free models in which the surface convective zones have solar metallicity and metal mass fraction of $Z = 10^{-4}$}\label{fig:conv}
\end{figure}

The computations were also made for the models of low-mass stars with zero- and low-metallicities. The model of $M = 0.8 \msun$ with the mass fraction, $Z$, of elements heavier than helium set at zero was constructed at the age of 10 Myr from the zero-age main sequence, which was used as a target star in the 3D simulations. Other models of $M = 0.82 \msun$ with various metallicities were computed to check the effect of surface pollution on the low-mass stars. In the following, we present the models of $0.8 \msun$ and $0.82 \msun$ stars in the same plot, but there are only minor quantitative differences in these models. Figure~\ref{fig:conv} shows the evolution of the mass of the surface convective zones from the main sequence phase to the beginning of the red giant phase.

To confirm the effect of surface pollution, we computed the models by covering metal-rich ($Z = 10^{-4}$ and $Z = Z_{\odot}$) materials in the surface convective zones. The depth of the convective zone is almost the same as the models without pollution because the structure of the surface convective zone is determined by the nuclear burning in the centre.

\subsection{Sample selection from the database}

\begin{figure*}
 \begin{center}
  \includegraphics[width=160mm]{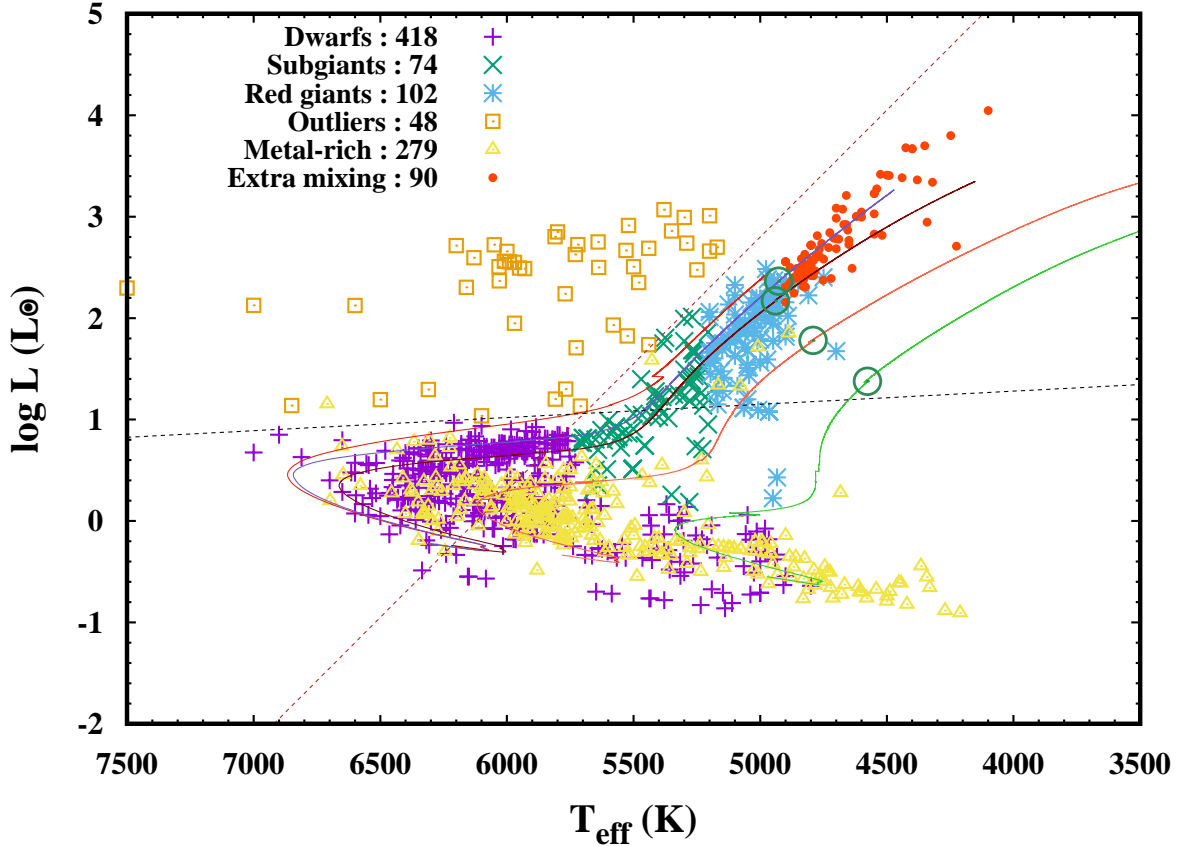}
 \end{center}
 \caption{Sample stars on the H-R Diagram. Luminosities are calculated from atmospheric temperatures and surface gravities in the original data by assuming that the stellar mass is $0.8 \msun$. The solid curves are stellar evolutionary tracks of $0.82 \msun$ with the metallicities of $\feoh = - \infty, -3, -2, -1$, and $0$ from left to right. The circles on the tracks denote the locations of the RGB bumps. The stars are divided into subclasses depending on their evolutionary stages such as dwarfs, subgiants, and red giants. The definition of the classifications is described in the text. We have excluded some stars with peculiar positions on the H-R diagram, as separated by the two dotted lines in the figure. The number of stars in each category is shown in the figure.}\label{fig:sample}
\end{figure*}

The data have been taken from the December 11, 2019 version of the Stellar Abundances for Galactic Archaeology (SAGA) database \citep{Suda2008,Suda2011,Yamada2013,Suda2017}.
The differences in the adopted solar abundances among different papers are properly taken into account for metallicity in figure~\ref{fig:life} \citep{Suda2017}.
Individual stars are separated by different evolutionary stages.
Dwarfs are defined as the effective temperature, $\teff > 5727$ K or the surface gravity, $\logg > 4.0$, based on the stellar model of $0.82 \msun$ with $\feoh = -3$.
The boundary between the subgiant and the red giant is determined by $\teff = 5200$ K.
There are some outliers on the H-R diagram, which are excluded from the sample. The boundaries of outliers are arbitrarily determined by the two dotted lines in figure~\ref{fig:sample}:
\begin{eqnarray}
\log L / L_{\odot} = -0.0025 \teff + 15.3 \\
\log L / L_{\odot} = -0.00013 \teff + 1.8
\end{eqnarray}
We have removed sample stars that are likely affected by extra mixing during the RGB phase.
There is increasing evidence of lithium depletion by extra mixing in red giants above the RGB bump \citep{Charbonnel2020}.
These stars are removed from the sample by defining the boundary of the RGB bump, which is estimated by the comparison of stellar models with the observed effective temperature, luminosity, and metallicity.
We identified the location of the RGB bump for $0.82 \msun$ with $\feoh = -3, -2, -1$, and 0 as shown in figure~\ref{fig:sample}.
If the observed stars have larger luminosities and lower effective temperatures than those of the RGB bump, they are classified as ``extra mixing'' stars.
The metallicity dependence of the boundary of the RGB bump is calculated by the interpolation or extrapolation of the above four data points.
In total, 90 stars out of 200 red giants with $\feoh < -1$ are classified as extra mixing stars.
As expected, the majority of extra mixing stars show lithium depletion after the correction with stellar evolution models (see figure~\ref{fig:life}).

\subsection{SPH simulations}

We conducted a series of hydrodynamical simulations of binary systems of the first stars, to estimate the evolution of the surface lithium abundances and metallicities.
In this section, we describe the models, numerical methods, and some comparison results, which are necessary to select the fiducial models.
Table~\ref{tab:model} provides the model parameters in this study.
The model H15F is the fiducial model in this study.
The reason for the choice of our numerical setups can be found in the following subsections and appendix.
The model names with ``S'' stand for ``separation'' to investigate the dependence on binary separation.
The model name with ``R'' stand for ``resolution'' to refer to resolution studies, and the model with ``SSPH'' corresponds to the investigation with the standard SPH method.
The model with ``W'', which means ``whole'', adopts the highest resolution in the whole simulation volume for the ejecta without using anisotropic particle-mass distribution, as described in \S~\ref{sec:distshell} and \ref{sec:distdirection}.
See appendix for our feasibility studies on resolution and computational methods, correspoding model names with ``R'', ``W'', and ``SSPH''.

\subsubsection{Numerical Simulation Code}

Numerical simulations shown here were carried out by a parallel code ASURA \citep{Saitoh2008,Saitoh2009}, which was originally developed for the simulations of galaxy formation.
The hydrodynamic equations are solved by a smoothed particle hydrodynamics (SPH) method \citep{Lucy1977,Gingold1977} with an improved version of the SPH (density-independent formulation of SPH; DISPH \citep{Saitoh2013}) implemented.
The conventional formulation of SPH assumes the smoothness of the density field and cannot handle fluid instabilities due to the unphysical surface tension that appears at contact discontinuities.
We employed the simple equation of state (EoS) that assumes ideal and mono-atomic gas (i.e., the specific heat rate is $\gamma = 5/3$).
Chemical reactions in the ejecta and the star are not taken into account.
Self-gravity is solved by the Tree method \citep{Barnes1986}.
We only considered the monopole moment with the tolerance parameter $\theta = 0.5$.
The interactions between particle-particle and particle-monopole moment are computed with Phantom-GRAPE \citep{Tanikawa2012,Tanikawa2013}.

\subsubsection{Setup of Initial Condition for SPH Simulations}

\begin{figure}
 \begin{center}
  \includegraphics[width=80mm]{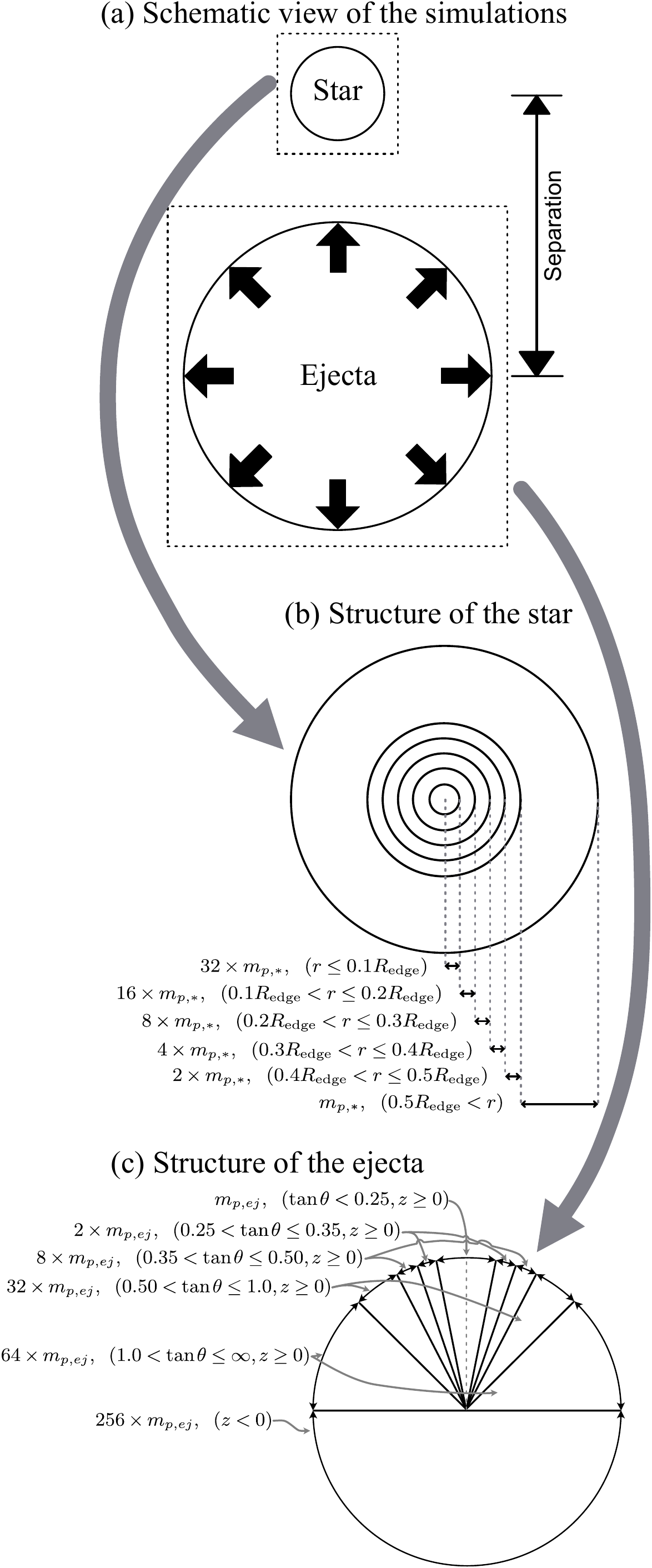}
 \end{center}
 \caption{Schematic pictures of the simulation setup. (a) The circle at the bottom represents the supernova ejecta after the final phase of the evolution of a massive star. The circle at the top is the low-mass star companion in the binary system. The arrows represent the ejection of matter from the massive star. The initial separation between the two stars is measured by the distance between the centre of the supernova ejecta and the low-mass companion. (b) The initial structure of the star. The inner region consists of coarse particles to reduce the computational cost. (c) The initial structure of the ejecta. The ejecta are divided by the direction so that the region has the finest resolution for colliding angles.}\label{fig:setup}
\end{figure}

Figure~\ref{fig:setup} shows the schematic picture of our model of a binary system.
It consists of a low-mass star and supernova ejecta.
One-dimensional models are mapped onto the particle distributions in a three-dimensional volume.
We fixed the mass of the low-mass star and three different supernova ejecta named H15, H20, and H25 which represents the 15, 20, and $25 \msun$, respectively, from the Heger's pre-supernova models \citep{Heger2010}.
We tested four different separations to investigate the dependence of the stripping and accretion of the surface of the low-mass star by the collision of ejecta.
The impact of the initial separation on the evolution is studied for the H15 model by adopting four different separations.
The separations of the H20 and H25 models are the minimum allowable in which the low-mass star and supernova progenitor contact with each other.
The minimum separation in the H15 model also has this configuration (See \S~\ref{sec:initial} for further details).
We assumed that the convective zone of the surface of the model mixes the materials homogeneously and does not change its mass and structure by the collision of supernova ejecta.
This assumption may affect the estimate of the final chemical composition by the simulations.
We ignored the rotation of a low-mass star with respect to the supernova progenitors because its timescale is much longer than those of the expanding ejecta.
In our simulation setups, binary components should be tidally synchronized so that the effect of rotation is not significant.

\subsubsection{Mapping the 1D Stellar Model onto the 3D space} \label{sec:distshell}

\begin{figure*}
 \begin{center}
  \includegraphics[width=160mm]{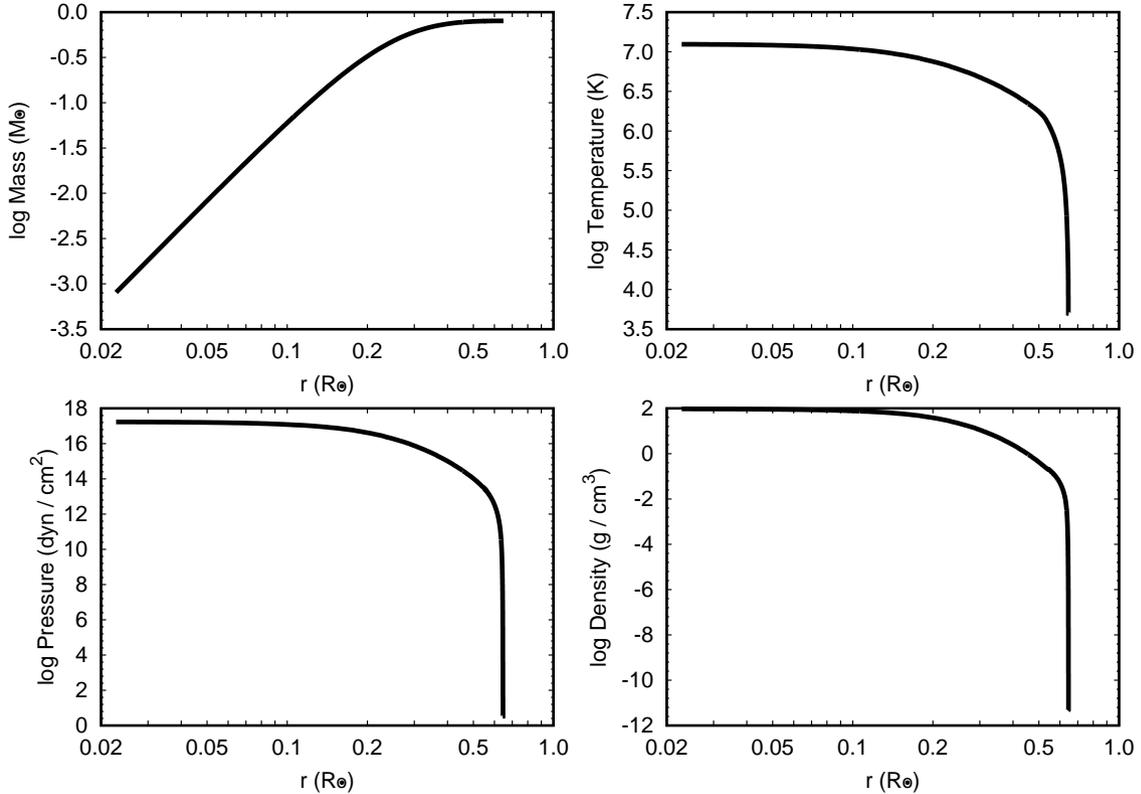}
 \end{center}
 \caption{Initial radial profiles of the low-mass star. The mass of the star (top left) denotes the integrated mass within radius r. The temperature (top right), pressure (bottom left), and density (bottom right) denote the values at the shell r.}\label{fig:profile}
\end{figure*}

We used the result of the numerical simulation of $0.8 \msun$ with $Z = 0$ as described above.
Figure~\ref{fig:profile} shows the radial profiles of the three-dimensional model star taken from the one-dimensional model at $t = 10$ Myr from the zero-age main sequence.

To construct a three-dimensional model with particles, we put the particles in individual shells so that the mass contained in ($r_{\rm in}, r_{\rm out}$) satisfies the condition,
\begin{equation}
\int_{r_{\rm in}}^{r_{\rm out}} 4 \pi r^{2} \rho \left( r \right) dr =m \left( r_{\rm in} \right),
\end{equation}
where $\rho (r)$ is the density profile and $m(r_{\rm in})$ is the particle mass at $r_{\rm in}$.
The exact position of a particle in the shell is determined by $R(r_{\rm out} - r_{\rm in}) + r_{\rm in}$ with a random number $R \in [0,1]$.
The values of the density at individual positions are given by linear interpolation between $\rho \left( r_{\rm in} \right)$ and $\rho \left( r_{\rm in} \right)$ and $\rho \left( r_{\rm out} \right)$.
The angular positions, the polar and azimuthal angles, are also randomly chosen.
The central particle is distributed by finding $r_{\rm out}$ from the above equation for $m \left( r_{\rm in} \right) = 0$ and $m \left( 0 \right) = 32 m_{p,*}$ where $m_{p,*}$ is the minimum mass of an SPH particle and is set at $3 \times 10^{-7} \msun$.
For the rest of the mass distribution $m \left( r \right)$, we adopted the following formula to reduce the computational cost,
\begin{eqnarray}
m \left( r \right) =
  \left\{
    \begin{array}{lr}
      m_{p,*}, & \left( 0.5 R_{\rm edge} < r \right) \\
      2 \times m_{p,*}, & \left( 0.4 R_{\rm edge} < r \leq 0.5 R_{\rm edge} \right) \\
      4 \times m_{p,*}, & \left( 0.3 R_{\rm edge} < r \leq 0.4 R_{\rm edge} \right) \\
      8 \times m_{p,*}, & \left( 0.2 R_{\rm edge} < r \leq 0.3 R_{\rm edge} \right) \\
      16 \times m_{p,*}, & \left( 0.1 R_{\rm edge} < r \leq 0.2 R_{\rm edge} \right) \\
      32 \times m_{p,*}, & \left( r \leq  0.1 R_{\rm edge} \right) \\
    \end{array}
  \right.
\end{eqnarray}
where $R_{\rm edge} \approx 0.64 \rsun$.
The outer layers have finer mass resolutions. The total number of gas particles is 1,023,991.
The star has the chemical composition of $X = 0.767$, $Y = 0.233$, $Z = 0$, where $X$, $Y$, and $Z$ represents the mass fraction of hydrogen, helium, and other elements, respectively, at the age of 10 Myrs from the onset of hydrogen burning in the centre.

The internal energy of each particle is computed by the linear interpolation of the temperature profile with the assumption of an ideal gas with the adiabatic index of $\gamma = 5/3$ and the mean molecular weight of 0.6.

Before starting the simulations of ejecta collisions, we relaxed the particle distribution of a three-dimensional model star following the simulations for Type Ia SNe \citep{Rimoldi2016}.
We added the dumping term in the hydrodynamical force and integrated the star for 2 hours, which is sufficient for the relaxation because the dynamical time of the system is estimated to be 30 minutes.
This technique enables us to avoid the oscillations of the stellar surface, which is not suitable for our simulations due to the change of the cross-section of collisional particles. 

\subsubsection{Mapping the 1D Ejecta Model onto the 3D space} \label{sec:distdirection}

\begin{figure*}
 \begin{center}
  \includegraphics[width=160mm]{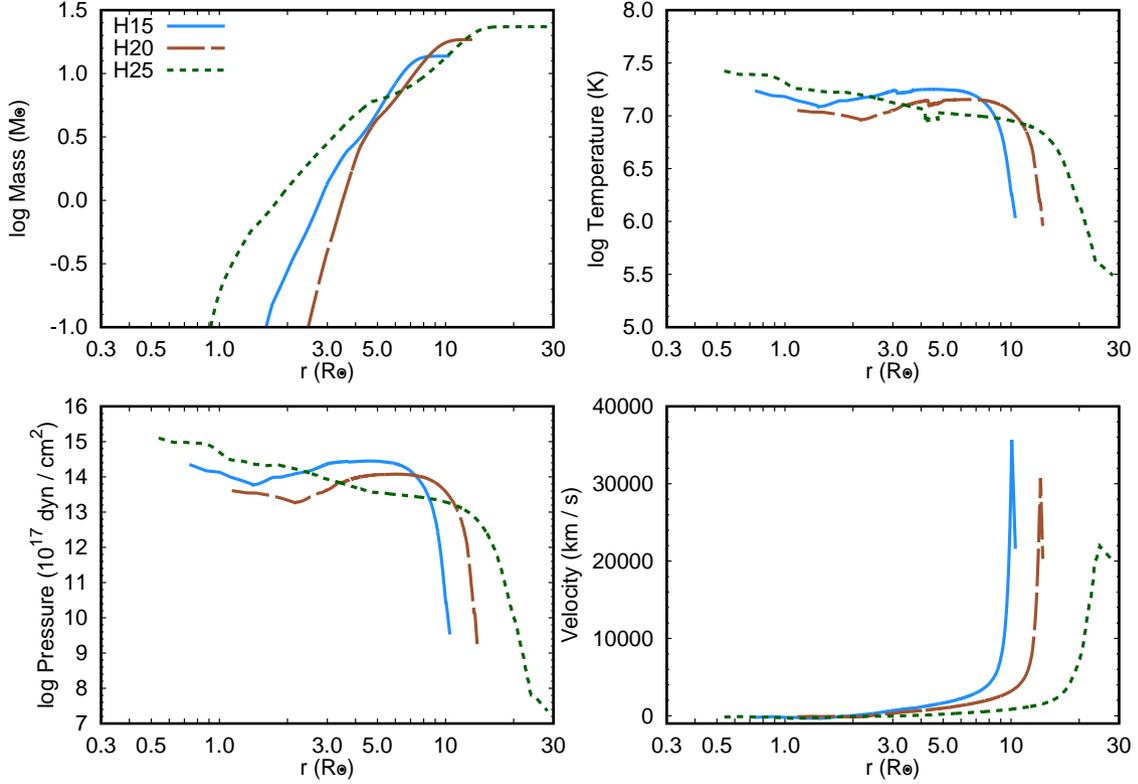}
 \end{center}
 \caption{Initial radial profiles of the three ejecta models. The comparisons of H15, H20, and H25 are made for the masses within the radii (top left), the temperatures (top right), pressures (bottom left), and velocities (bottom right) of the ejecta.}\label{fig:ejecta}
\end{figure*}

The one-dimensional profiles of pre-supernova models with $Z = 0$ are taken from the literature \citep{Heger2010}.
We have employed the 15, 20, and $25 \msun$ models.
Figure~\ref{fig:ejecta} shows the radial profiles of the three supernova ejecta models.
The outer edge of ejecta is 10, 14, and $24 \rsun$, and the corresponding time from ignition is 1747 sec, 2195 sec, and 5705 sec, respectively.
The model of $15 \msun$ is prescribed in the model of SN 1987A \citep{Shigeyama1990}.

We have developed a new technique to ensure a high resolution in colliding particles by changing the particle mass depending on the direction of the ejecta particles.
Here we define the direction of the ejecta particles by the angle $\theta$ from the centre of the ejecta along the $z$ axis connecting with the centre of the companion star.
The particle mass depends on the initial location of the ejecta on the ($\theta, z$) coordinate as follows,
\begin{eqnarray}
m \left( \theta, z \right) =
  \left\{
    \begin{array}{lr}
      m_{p,ej}, & \left( \tan \theta < 0.25, z \geq 0 \right) \\
      2 \times m_{p,ej}, & \left( 0.25 < \tan \theta \leq 0.35, z \geq 0 \right) \\
      8 \times m_{p,ej}, & \left( 0.35 < \tan \theta \leq 0.50, z \geq 0 \right) \\
      32 \times m_{p,ej}, & \left( 0.50 < \tan \theta \leq 1.0, z \geq 0 \right) \\
      64 \times m_{p,ej}, & \left( 1.0 < \tan \theta \leq \infty, z \geq 0 \right) \\
      256 \times m_{p,ej}, & \left( z < 0 \right) \\
    \end{array}
  \right.
\end{eqnarray}
where $m_{\rm p,ej}$ is the minimum mass of an ejecta particle.
The positions of the particles are determined in the same way as described in the previous subsection. The radial velocities, internal energies, and densities for individual particles are assigned to reproduce the profile in figure~\ref{fig:ejecta}.
We tested four different choices of $m_{\rm p,ej}$, i.e., $\pow{1}{-6} \msun, \pow{3}{-7} \msun, \pow{1}{-7} \msun$, and $\pow{3}{-8} \msun$.
The third choice (H15) is adopted as the fiducial resolution because the result converges with the result with the highest mass resolution (H15Rc).
The corresponding numbers of particles for the four models are 479,442, 1,599,777, 4,800,569, and 16,001,929, respectively.
Those for H20 and H25, with the fiducial mass resolution, are 6,487,398 and 8,189,647, respectively.

We have confirmed the validity of using particles with different masses by running a test simulation using equal-mass particles and compared it with the direction-dependent mass model for the same condition for the collision.
It is confirmed that the stripped mass from the companion star and the accreted mass are consistent with each other.
We found that we need $\approx$ 140,000,000 particles for the equal-mass model to achieve a similar number of effective particles ($\tan \theta < 0.25, z \geq 0$), i.e., our new technique is successful in reducing the number of particles by $1 / 30$.

Thanks to our improved method for the assignment of mass in the SPH particles, more than 130,000 particles effectively interact with the companion star despite the small visual angle of 3.5 degrees in the H15 model.

The consistency between 1D and 3D models is confirmed by the total kinetic energy just before the impact of the ejecta.
We also checked the stripped mass of the companion with the setup of Type Ia supernovae and compared it with the previous study under the same initial mass and separation \citep{Pakmor2008}.

\subsubsection{Initial conditions for a binary system}\label{sec:initial}

In the fiducial model, we adopt the initial separation of 0.1 au ($21.5 \rsun$), which is larger than the radius at supernova explosion according to metal-free star models \citep{Heger2010} and the critical radius of the Roche-lobe overflow based on the empirical formula \citep{Eggleton1983}.
We also computed the cases with 0.2, 0.4, and 0.8 au to investigate the dependence on the initial separation.
To compare various progenitors, we computed the models of 15, 20, and $25 \msun$ stars with the minimum separations that are characterized by the size of the progenitor stars.
These correspond to 0.048 au, 0.064 au, and 0.112 au, respectively.

The orbital motion of the binary system was ignored in this study for the sake of simplicity.
This is reasonably validated by the short timescale ($\sim 10$ hrs) in our simulations compared with the orbital period (a few days to a few months).
The consideration of the orbital motion may have some effect on the amount of accretion at the separation of 0.1 au, while the stripping of the surface layers will not be affected because the stripping is dominated by massive fast ejecta in a shorter timescale (a few hours).
It is worth investigating the case of accretion with the motion of a binary orbit taken into account, but it is beyond the scope of this paper.

By ignoring the centrifugal force by the orbital motion, a gravitational pull exerts on the ejecta and the star, which results in an unwanted motion in the system.
Therefore, we filtered a long-range force of gravity for the companion star by ignoring the gravitational interaction with the ejecta at the distance of more than $10 \rsun$ from the centre of the star.
This prescription mimics the orbital motion and is sufficient to avoid the strong gravity caused by the ejecta at the initial position and to consider the effective accretion process after the collision.

Mass loss from the progenitor stars are not considered, which is justified by the metallicity dependence of stellar winds. 

The rotation of the star and the ejecta is not considered.
The consideration of rotation will be more realistic, especially for the ejecta, but it is too complicated to implement in the simulations.
On the other hand, fast rotation in companion stars will not be mandatory in the binary systems studied.

We also employed a particle-split method \citep{Kitsionas2002,Martel2006} to retain a high resolution even in large separations.
In the run of the H15Sa model, the ejecta particles within $6 \rsun$ from the centre of the low-mass star split into 8 smaller particles.
In the runs of the H15Sb and H15Sc models, those within $12 \rsun$ split into 8 smaller particles and those smaller particles divided into 8 further smaller particles when they are within $6 \rsun$.

\begin{table*}
  \caption{Simulated models and results. The column represents the model name, the number of SPH particles for a companion star and supernova ejecta, the binary separation, the unbound mass by the stripping, the bound mass by the accretion, the final iron and lithium abundance without and with lithium-rich ejecta in the surface of the companion star. See \S~\ref{sec:discussion} for the details of estimating the abundances.}\label{tab:model}
  \begin{center}
    \begin{tabular}{l*{8}{r}}
    \hline
Model & $N_{\rm star}$ & $N_{\rm ejecta}$ & Separation & Stripped mass & Accreted mass & $\feoh_{\rm mixed}$ & $\ali_{\rm woLi}$ & $\ali_{\rm wLi}$ \\
  &  &  &  au  &  $10^{-3}$ \msun & $10^{-3}$ \msun &  &  & \\
    \hline
H15 & 1,023,991 & 4,800,569 & 0.048 & 2.70 & 3.21 & 0.43 & 1.24 & 4.02  \\
H20 & 1,023,991 & 6,487,398 & 0.064 & 2.08 & 4.13 & 0.54 & $-\infty$ & 4.41  \\
H25 & 1,023,991 & 8,189,647 & 0.112 & 1.06 & 2.20 & 0.05 & 1.71 & 4.14  \\
H15F & 1,023,991 & 4,800,569 & 0.1 & 1.43 & 0.713 & -0.22 & 2.01 & 2.79  \\
H15Sa & 1,023,991 & 4,800,569 & 0.2 & 0.700 & 0.0972 & -1.09 & 2.09 & 2.12  \\
H15Sb & 1,023,991 & 4,800,569 & 0.4 & 0.268 & 0.0145 & -1.92 & 2.10 & 2.10  \\
H15Sc & 1,023,991 & 4,800,569 & 0.8 & 0.216 & 0 & $-\infty$ & 2.10 & 2.10  \\
H15Ra & 1,023,991 & 479,442 & 0.1 & 3.35 & 0.826 & -0.16 & 1.99 & 2.90  \\
H15Rb & 1,023,991 & 1,599,777 & 0.1 & 2.03 & 0.813 & -0.17 & 1.99 & 2.88  \\
H15Rc & 1,023,991 & 16,001,929 & 0.1 & 1.32 & 0.76 & -0.20 & 2.00 & 2.83  \\
H15W & 1,023,991 & 137,215,327 & 0.1 & 1.29 & N/A* & N/A & N/A & N/A  \\
H15SSPH & 1,023,991 & 4,800,569 & 0.1 & 0.899 & 0.908 & -0.12 & 1.98 & 2.97 \\
     \hline
     \end{tabular}
  \end{center}
\end{table*}

\section{Results}\label{sec:results}

\begin{figure*}
 \begin{center}
  \includegraphics[width=180mm]{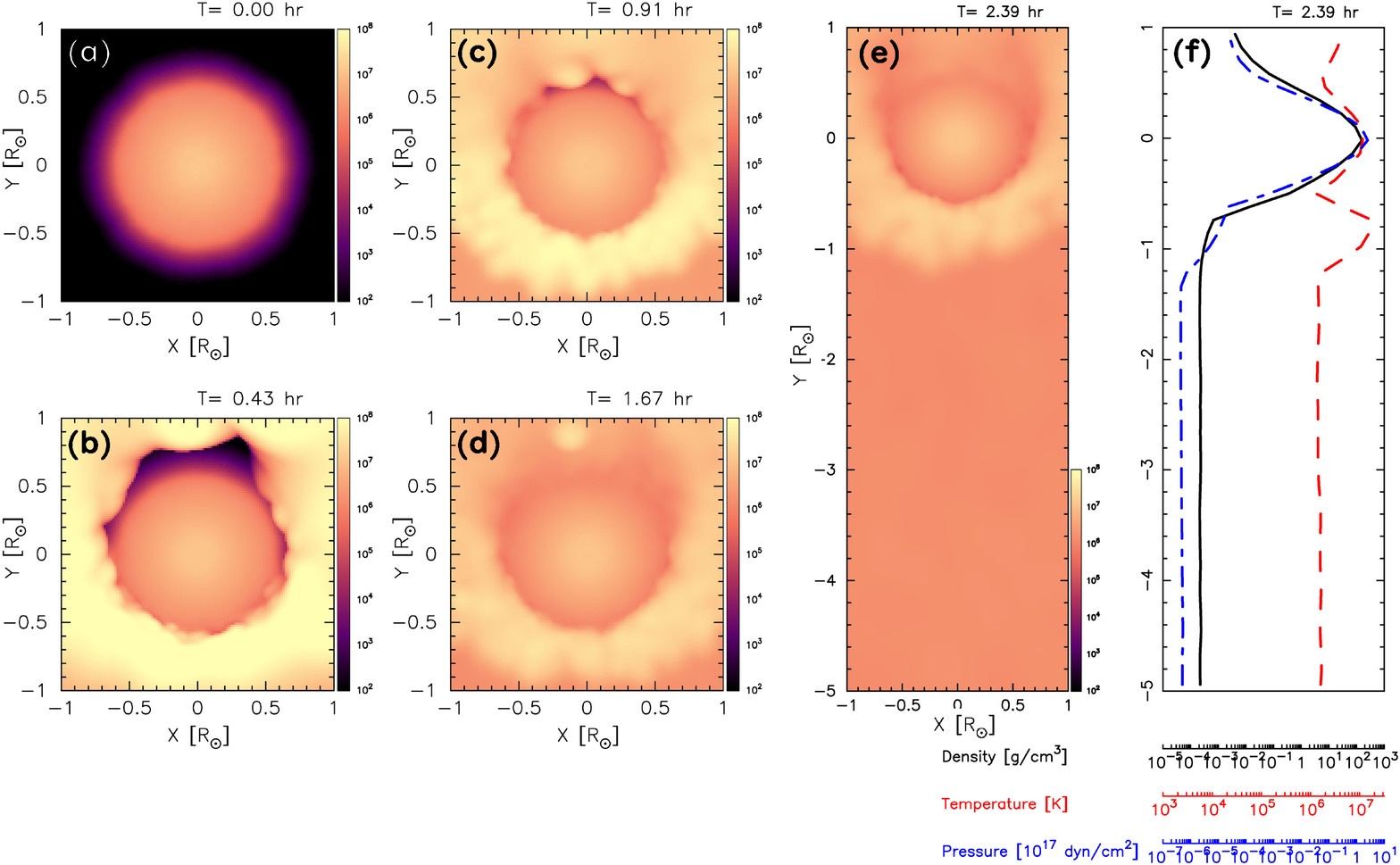}
 \end{center}
 \caption{Simulation snapshots of the H15 model.
 Panels (a), (b), (c), (d) and (e) represent the temperature distribution around the target star at the times of 0.0, 0.43, 0.91, 1.67, and 2.39 hr, respectively.
 The target star with $0.8 \msun$ is located on the top as shown in the panel (e).
 The blast wave of a supernova comes from the bottom.
 Panel (f) shows the profiles of density, temperature and pressure of the gas along the axis which connect the centres of the target star and ejecta.}\label{fig:sph}
\end{figure*}

Figure~\ref{fig:sph} shows the result of simulations for a binary system consisting of 15 \msun\ and 0.8 \msun\ stars separated by 0.048 au (the H15 model in table~\ref{tab:model}).
The impact of the ejecta onto the binary companion is simulated for 24 hours after the explosion for this model.
Once the ejecta material collides with the surface of the companion star, a bow shock is formed near the surface.
The basic picture of the outcome of the collision is consistent with previous studies \citep{Pakmor2008,Liu2013,Hirai2018}.

\begin{figure}
 \begin{center}
  \includegraphics[width=80mm]{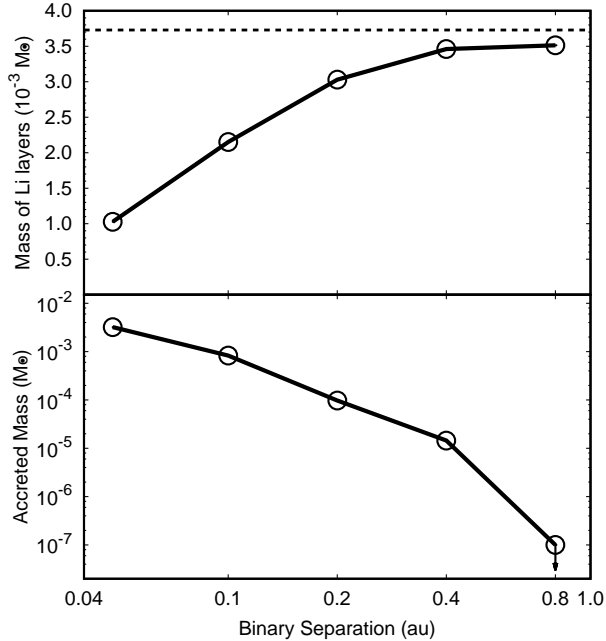}
 \end{center}
 \caption{Mass of lithium-containing layer due to the stripping by the ejecta (top panel) and accreted mass (bottom panel) for the 15 \msun\ models. The values are taken from the models H15, H15F, H15Sa, H15Sb, and H15Sc from left to right. The dotted line denotes the mass of the surface convective zone for the 0.82 \msun\ model.}\label{fig:mass}
\end{figure}

There are two possible impacts on the surface of a binary companion, induced by the collision of supernova ejecta.
Fast-moving outer ejecta, having a velocity as much as 10,000 km/s, strip the surface layers of the companion star, while a part of slowly moving inner ejecta, down to 3,000 km/s, are accreted by the companion.
It is found that the ejecta do not strip the whole of the surface convective zone.
Therefore, the surface chemical composition does not change due to this effect (top panel in figure~\ref{fig:mass}).
The mass of the accreted matter is estimated by calculating the total energy of individual particles representing the ejecta (bottom panel in figure~\ref{fig:mass}).

We found that the particles near the surface of the companion stars experienced a shock heating that gave rise to high temperatures with $\gtrsim 10^{8}$ K.
At this temperature, \nucm{7}{Li} can be destroyed quickly even for this short timescale event.
The density of the shocked region is the order of $10^{-4} {\rm g}~{\rm cm}^{-3}$, which is lower than the density, $\sim 1 {\rm g}~{\rm cm}^{-3}$, of the shell where \nucm{7}{Li} burns in the envelope with $T \approx \pow{3}{6}$ K.
On the other hand, the nuclear reaction rates are much larger at $10^{8}$ K than at $\pow{3}{6}$ K by 16 orders of magnitude \citep{Caughlan1988}.
The nuclear timescale for lithium destruction by proton-capture reactions is estimated to be $\sim 300$ sec.
However, the particles affected by shock heating do not accrete onto the surface of the companion star and escape from the system.
Therefore it is justified to ignore the change of the surface chemical composition by nuclear reactions in our hydrodynamical simulations.

\begin{figure}
 \begin{center}
  \includegraphics[width=80mm]{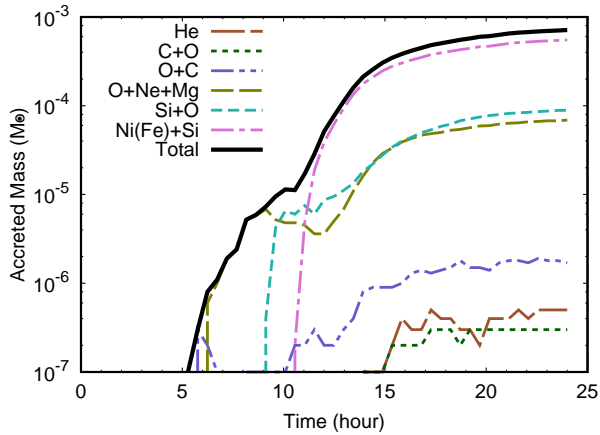}
 \end{center}
 \caption{Temporal change of the mass of gravitationally bound particles around the low-mass star for the model H15F. The accreted mass of the ejecta is separated by the chemical composition of the supernova ejecta, under the assumption that the supernova ejecta are not mixed.
The line for the hydrogen-rich envelope is not shown because no SPH particles accreted on the surface.
 }\label{fig:yield}
\end{figure}

Figure~\ref{fig:yield} shows the time evolution of the mass of bound ejecta particles grouped by chemical compositions.
The inner ejecta, such as the nickel- and silicon-rich layers, result in more efficient accretion than the outer ejecta like the hydrogen- and helium-rich layers.
This can be interpreted as strong velocity dependence of the accretion rate, where the amount of accretion is larger for slower ejecta.
If the ejecta shells were unmixed and retain their chemical composition, the ejecta shells within the oxygen, neon, and magnesium layers would be accreted onto the companion.
This means that the final surface abundances of companion stars would be very metal-rich with unusual abundance patterns.
However, supernova ejecta are thought to be well mixed during the shock propagation in the envelope by the Rayleigh-Taylor instability, inferred from the timing of the detection of line gamma rays and X-rays in SN 1987A \citep{Arnett1989,Kumagai1989}.
It is to be noted that there are multi-dimensional simulations of the SN 1987A explosions \citep{Utrobin2019}, but the yields of supernova ejecta are not available from their results.
Although our models are not able to predict the abundance patterns, it will be interesting to specify the progenitors by characteristic abundance patterns in the first supernovae.
If the observed stars reflect abundance patterns by core-collapse supernovae, we may observe deficiencies of odd-Z elements.
Also, the abundance ratios of [Na/Mg] or [Ca/Mg] can be diagnoses for pair-instability supernovae \citep{Takahashi2018}, which can be used to discriminate our scenario from the second generation stars formed out of the ejecta of pair-instability supernovae.

The accretion process is subject to uncertainties in estimating the bound mass.
We estimated the expected amount of the bound mass by considering the balance between the gravitational binding energy and the kinetic energy of the individual particles of the ejecta.
This is because the accretion process is complicated and the phenomenon takes a much longer timescale than the time of the simulations.
To estimate accurate gravitational force exerted on the individual particles, we iterated several times the effect of gravity after removing the particles that are going to escape from the system.

It is highly uncertain how the accreted gas particles mix with the convective envelope of the companion star.
In our simulations, we do not follow the accretion process, which proceeds in a thermal timescale, due to a much longer timescale than the dynamical process in this study, and the technical difficulty to simulate gas accretion onto a stellar surface.
The detailed modelling of the accretion process is poorly understood due to the complexity of the formation of accretion discs and the considerations of radiation pressure, magnetic force and other physical processes on a stellar surface \citep{Marietta2000}.
Still to be established is how the convective envelope is reconstructed after the gas accretion.
There are no adequate prescriptions to implement the mixing of matter into the convective envelopes either in 3D hydrodynamical simulations or in a 1D hydrostatic stellar evolution code.

To consider the effect of mixing suggested by observations as above, we assumed that the bound ejecta particles are well mixed in the convective envelope of companion stars.
We obtained $\feoh = -0.2$ for well-mixed ejecta in the case of model H15 where the maximum amount of accretion is achieved.
The lithium abundance will be reduced by 0.86 dex, from the estimate of the accreted mass and the mass of the surface convective zone for a metal-free 0.82 \msun\ model at the time of impact, when $10^{7}$ years have passed since the onset of hydrogen burning.
For longer separations, lithium depletion is much less.
In the $15 \msun$ case, lithium depletion is negligible at 0.2 au.
In larger separations, the amounts of stripping and accretion decrease very rapidly.
The results for all the models are summarized in table~\ref{tab:model}.
The lithium abundances will not change after the accretion event since the surface convective zone of low-mass stars recedes in mass during the main sequence evolution (figure~\ref{fig:conv}) and will not mix with the lithium-containing layer.

\section{Discussions}\label{sec:discussion}

The values of $\feoh_{\rm mixed}$ in table~\ref{tab:model} are calculated by the following equation.
\begin{equation}
\feoh_{\rm mixed} = \log \frac{ M_{\rm acc} }{ M_{\rm ini} - M_{\rm rem} } Y_{\rm Fe} - \log X_{\rm H} M_{\rm conv} - \log \frac{ X_{\rm Fe, \odot} }{ X_{\rm H, \odot} },
\end{equation}
where $M_{\rm acc}$ denotes the accreted mass taken from the simulation results in table~\ref{tab:model}.
Other parameters are taken from stellar models: $M_{\rm ini}$, $M_{\rm rem}$, and $M_{\rm conv}$ is the initial mass, the remnant mass which is set at $1.3 \msun$, and the mass of the surface convective zone of the low-mass companion which is set at $\pow{3.73}{-3} \msun$, respectively.
The iron yields from supernovae, $Y_{\rm Fe}$, are taken from the mixed models with the explosion energy of \pow{1.2}{51} erg in table 6 of \citet{Heger2010}.
The abundance parameters, $X_{\rm Fe, \odot}$ and $X_{\rm H, \odot}$, are taken from the literature \citep{Asplund2009}.
The hydrogen abundance of $X_{\rm H}$ in the surface of low-mass first stars are assumed to be the solar value.

The final lithium abundances are estimated by the following equation.
\begin{eqnarray}
\ali &=& \log \left[ \frac{ M ( {\rm Li} ) }{ M_{\rm conv} } \frac{ 1 }{ A_{\rm Li} X_{\rm H, \odot} } \right] + 12, \nonumber \\
     &=& \log \left[ \left( \frac{ M_{\rm acc} }{ M_{\rm conv} } \right)^{2} \frac{ Y_{\rm Li} }{ M_{\rm ej} } \frac{ 1 }{ A_{\rm Li} X_{\rm H, \odot} } f \right. \nonumber \\ 
     & & \hspace{1.5cm} \left. + 10^{A_{\rm Li, ini} - 12} \frac{ M_{\rm conv} - M_{\rm acc} }{ M_{\rm conv} } \right] 
      + 12,
\end{eqnarray}
where $M( {\rm Li} )$, $A_{\rm Li}$, $Y_{\rm Li}$, $M_{\rm ej}$, and $A_{\rm Li,ini}$ denote the lithium mass in the surface convective zone, the mass number of $^{7}{\rm Li}$, lithium yields from a supernova, the total mass of supernova ejecta, and the initial lithium abundance, respectively.
The values of $Y_{\rm Li}$ and $M_{\rm ej}$ are taken from the same models as above.
The initial value of lithium abundance is assumed to be 2.1.
The free parameter $f$ depicts the contribution of the lithium yield in the supernova ejecta.
In table~\ref{tab:model}, we provided the final lithium abundances without lithium yields, $A( {\rm Li} )_{\rm woLi}$, by the above equation by setting $f = 0$ and those with lithium yields, $A( {\rm Li} )_{\rm wLi}$, by setting $f = 1$.

It is to be noted that the above two equations do not include the stripped mass.
This is because the mass of lithium-containing layers is always smaller than the mass of the surface convective zone of the low-mass companion in our simulations.
For instance, the interaction between the ejecta and the companion can strip the envelope with a mass of $\pow{3.3}{-3} \msun$ at most as shown in figure~\ref{fig:mass} and table~\ref{tab:model}.
The stripped mass is smaller than the mass of the surface convective zone of a $0.82 \msun$ star at the age of $\sim 10$ Myr, which is the lifetime of a $15 \msun$ star.
In this case, the stripping event will not reduce the surface lithium abundances, although it depends on how we treat the mixing in the surface convective zones, as there are no reliable models and theories on it.
Here we expect that the surface convective zones will be reconstructed after the accretion event rather than the stripping event.
The duration between the stripping and accretion is much shorter than the time for the surface convective zone to reconstruct.
Therefore, the accretion events of ejecta overwrite the effect of the stripping.

Our models predict solar metallicity stars in the halo of our Galaxy for shorter binary separations.
Figure~\ref{fig:life} shows the comaprisons of lithium abundances.
Provided that the first massive stars explode with small radii (figure~\ref{fig:hrd}), it is more likely that solar metallicity stars in the halo are survivors of the first stars in the universe, originally having a chemical composition as a result of big-bang nucleosynthesis.
In the case of longer binary separations, our models may account for metal-poor ($\feoh \sim -3$) lithium-depleted stars and lithium-rich ($\ali > 3$) stars.
Metal-poor ($\feoh < -5$) stars without lithium-depletion are also possible, although we cannot distinguish between our binary scenario and the case with single stars.
The abundance patterns of all the measured elements should be checked individually, but they will vary according to the abundances of yields and the degree of mixing in the ejecta.
The observed lithium depletion around $\feoh = -3$ is not explained by our models due to a small amount of lithium depletion for longer separations and too much accretion for shorter separations.
Instead, our models predict a small or negligible depletion at $\feoh < -5$ and the existence of solar metallicity stars in the halo stars in our Galaxy.

To confirm the existence of the first metal-rich stars in the Galactic halo, we need to exclude the possibility of normal metal-rich halo stars.
In our proposed scenario, we have excluded the possibility of stars that survived the common-envelope phase because a companion star in the common envelope phase will accrete sufficient mass to become much more massive than $0.8 \msun$.
Such stars can be easily excluded from the position on the H-R diagram.
Even if there is an Wolf-Rayet star with a low-mass companion, its supernova ejecta must be too fast to be accreted onto its companion due to the absence of the hydrogen-rich envelope in the progenitor.
There is also an argument that the evolutionary path of the SN 1987A progenitor is not established.
The most promising scenario for the progenitor is that it is a binary merger (e.g. \cite{Morris2007}).
If this is the case, we need a third star to make our scenario work.
This is obviously an unusual case and such stars cannot be a major noise for the search for metal-rich halo stars.

The formation of solar metallicity stars in metal-free or metal-poor host halos could be another exceptional case to be considered.
According to the hydrodynamical simulations of star formation triggered by supernovae with initial conditions taken from cosmological simulations, second-generation stars after first supernovae have metallicities up to $\feoh \sim -1$ \citep{Wise2012,Smith2015,deBennassuti2017,Chiaki2020}.
It is not realistic to form solar metallicity stars with a single supernova in a metal-free or metal-poor cloud in its host cloud.
However, it may be possible if the kinetic energy of supernovae is very low and the swept-up mass is correspondingly small.
It would be intriguing to compare the abundances of second generation stars formed from low-energy supernovae with those of binary companion stars in close massive binaries.

There is a possiblitiy of the formation of solar metallicity stars in chemical evolution in the progenitors of the Galactic halo.
Simulations of metal-enrichment in the Galactic halo by early generation stars predict metallicity of up to $\feoh \sim -0.5$ \citep{Komiya2014,Sarmento2017,Cote2018}.
For example, \citet{Cote2018} found star-forming gas having solar-metallicities at the redshift of $z = 7.29$ by their simulations of the formation of the most massive galaxy realised by \citet{Wise2012}.
\citet{Sarmento2017} found a highly-polluted region with gas metallicity of $Z = 0.1 Z_{\odot}$ close to the centre of its host halo by their simulations up to $z = 5$.
These results may imply that the current chemical evolution models do not focus on the formation of the most metal-rich stars in the Galactic halo, i.e., we presume that solar metallicity stars would not form or survive in the halo.
If metal-rich stars are formed in the halo by consecutive star formation, their abundance patterns should be dominated by type II supernovae, not by type Ia supernovae.
We will be able to identify the progenitors of such stars by larger [$\alpha$/Fe] compared with disc stars with the same metallicity, if they exist.

It is difficult to follow the formation of close binary systems at metal-free environment using hydrodynamical simulations.
Since our scenario focuses on binaries with the separation less than $\sim 1$ au, we need simulations with very high-resolution. 
All of the current simulations of first star binaries deal with protostar binaries much wider than 1 au (see e.g. \cite{Sugimura2020,Susa2019}).
It is not clear if the formation of first star binaries with small separations are supported by numerical simulations.
However, it is interesting that higher-resolution studies found more possible close binaries by decreasing the minimum spatial resolution of the simulations from 20 au \citep{Stacy2013} to 5 au \citep{Stacy2016}.

We anticipate more applications of our scenarios to the origin of known peculiar stars, namely lithium-enhanced stars.
There is an argument about the production of lithium in the hydrogen-rich envelope of massive stars by neutrino processes \citep{Heger2010}.
If we simply apply the well-mixed lithium yield to the accretion onto the low-mass companion, the maximum surface lithium abundance can be $\ali \sim 3.0$.
This may account for some of the extremely lithium-enhanced stars as large as $\ali \sim 4.0$ among the main sequence stars \citep{Li2018}.
Such lithium-enhanced stars are very rare, comprising 0.1 percent of the total population in almost all the metallicity range \citep{Kirby2016}.
The fact implies that the ejection of lithium-rich ejecta and the accretion onto the companion involve a complicated physics.

Theoretical models are not able to explain lithium enhancement in the surface of low-mass stars.
The transportation of \nucm{7}{Be} and the decay to \nucm{7}{Li} in the envelope require high temperature at the bottom of the convective envelope, and hence only AGB stars with $M \gtrsim 4 \msun$ can be $\ali \gtrsim 3.0$ by self-enrichment \citep{Karakas2016}.
Low-mass red giants are thought to suffer from extra mixing at the red giant branch bump (see, e.g., \cite{Charbonnel2020}) and experience strong lithium-depletion.
Even if the parameter for extra mixing is adjusted, it is too difficult to explain lithium-rich giants \citep{Lattanzio2015}.
Nova outbursts produce a large amount of lithium \citep{Starrfield2020}.
However, there are no known mechanisms to form low-mass stars in the Galactic halo from yields from novae.
If there exist metal-rich and lithium-rich halo stars whose ingredients are dominated by nova yields, we may distinguish them by the abundance ratios of CNO isotopes \citep{Jose1998}.
Lithium enhancement in super-solar metallicity is also controversal.
Theoretical models predict that lithium abundances are enhanced at super-solar metallicity, while the observations reveal a decreasing trend of lithium abundances.
According to the models by \citet{Karakas2016}, AGB stars with $M \geq 4.75 \msun$ and $Z = 0.03$ evolve to lithium-rich stars up to $\ali \sim 5$, while the observations show the decrease in \ali\ at $\feoh > 0$ \citep{Guiglion2016}.
Chemical evolution models for lithium in the Galactic disc can reproduce lithium abundances as high as $\ali \approx 3.5$ \citep{Prantzos2017}.
Therefore, the only possible contaminants are metal-rich stars born in the Galactic disc that fly out into the halo.

An exploration of different set of parameters and assumptions could be of interest.
For instance, an assumption of a spherically symmetric explosion is not guaranteed.
Aspherical explosions or jet explosions will change the amount of stripping and accretion \citep{Tominaga2009}, while there is no convincing theory and observations for the geometry of supernova explosions.
Still, there will be a way to produce metal-rich survivors with these models.
A more eccentric orbit will have a variation in the stripping and accretion depending on the orbital phase at the explosion, although this effect can be incorporated into the binary separation.
It is difficult, and beyond the scope of this study, to estimate the overall uncertainties and these new parameters.
Various situations and models may better explain the diversity of the chemical compositions of known extremely metal-poor stars.

After the supernova event of the massive star, the binary system will be disrupted by the ejection.
The remnant of the supernova will have a velocity of a few 100 km/s \citep{Cordes1993}.
In the circumstance of the first binary star formation, it may be difficult for the remnants of the supernovae to be confined in the gravitational potential of the low-mass host halo.
The low-mass companion star can remain in its host halo if the orbital velocity is lower or comparable to the velocity dispersion of the host halo.

Currently, there is no direct observational evidence of metal-free and metal-poor binary systems that we proposed because they are too old for massive primaries to survive.
Therefore, search for observational counterparts must be made for nearby, metal-rich stars.
We searched for low-mass companions in OB star binaries \citep{Moritani2018}.
The analysis of the 10 target stars is still ongoing, among which HD 164438 is reported to have an intermediate-mass star with the mass ratio of $q = M_{2} / M_{1} = 0.1 \hyp 0.2$ \citep{Mayer2017}, where $M_{1}$ and $M_{2}$ is the initial mass of a primary and secondary, respectively.
Although we have not yet established the population of binary systems in the solar vicinity, we can expect some fraction of stars that are metal-rich counterparts of the first-generation massive binaries.
There are no reliable models and observations of binary star formation for the entire range of metallicity for OB stars, and hence we encourage the exploration of the binary star formation.

To explore the possibility of observational counterparts in the solar vicinity, we have compiled a catalogue of OB stars from the literature to find binaries where the masses of a primary star and a secondary star is $10 \hyp 20 \msun$ and less than $1 \msun$, respectively, with the separation less than 1 au.
This criterion is translated into the mass ratio $q < 0.1$ and the binary period of a few days to a few months.
We investigated more than 20 binary catalogues and surveys to compile the existing observational data.
These data include spectroscopic binaries \citep{Pourbaix2004}, eclipse binaries \citep{Malkov2006}, astrometry binaries \citep{Mason2001} and visual binaries (Sixth Catalog of Orbits of Visual Binary\footnote{http://www.usno.navy.mil/USNO/astrometry/optical-IR-prod/wds/orb6
}).
We have collected binary parameters for the stars classified as Galactic OB stars using the online catalogue \citep{Skiff2014}. 

\begin{figure}
 \begin{center}
  \includegraphics[width=80mm]{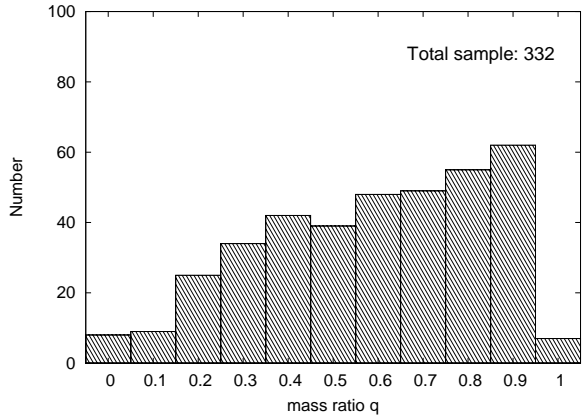}
 \end{center}
 \caption{Mass ratio distribution function from the literature data for the binaries whose period are 1year and shorter.}\label{fig:mratio}
\end{figure}

Figure~\ref{fig:mratio} shows the distribution of mass ratio taken from our compilation of 332 OB stars with known binary mass ratios and periods (1 year or shorter) from the observations.
There is an observational bias in the distributions of both periods and mass ratios in our sample.
In particular, binaries with small mass ratios are difficult to identify, and hence the number of such binaries is underestimated.
In the figure, no corrections for the distribution are considered as discussed in other studies \citep{Moe2017}.

Apart from the literature study, we have conducted a radial velocity monitoring to specify binary systems that are the metal-rich counterpart of those relevant to the massive binary scenario.
Medium-resolution spectra have been obtained with Medium And Low-dispersion Long-slit Spectrograph (MALLS) \citep{Ozaki2005} equipped to the Nayuta 2.0 m telescope at the Nishi Harima Astronomical Observatory (NHAO).
Also, high-resolution spectra has been obtained using two spectrographs, High Dispersion Echelle Spectrograph (HIDES) \citep{Izumiura1999,Kambe2013} equipped to the 188 cm telescope at the Subaru Telescope Okayama Branch Office (OBO), and Gunma Astronomical Observatory Echelle Spectrograph (GAOES) \citep{Hashimoto2006} equipped to the 1.5 m telescope at the Gunma Astronomical Observatory (GAO).
The details of the observations and analyses are on-going and an initial report is given in a separate paper \citep{Moritani2018}.

Candidates for the survivors of the first stars exist in our Galaxy.
A massive binary scenario presented here predicts the existence of metal-rich ($\feoh \sim 0$) and lithium-depleted, or perhaps lithium-rich ($\ali \sim 3$) stars in the Galactic halo.
We find 78 stars with $\feoh > -1$ and 6 stars with $\feoh > -0.5$ having the space velocities of halo stars in the SAGA database, although none of them have reported lithium abundances (table~\ref{tab:kin}).
Among them, one star, BS 17587-021 is apparently a metal-rich halo star from its metallicity ($\feoh = 0.93$, \cite{Lai2008}) and its motion different from the Galactic disc components, estimated using the data from the Gaia Data Release 2 (DR2) \citep{GaiaCollaboration2018}.
We need a more careful inspection of this star because only one iron line at $670.015$ nm was used to determine its metallicity.
There are also a number of solar metallicity stars among kinematically selected halo stars to search for high-velocity stars ejected from the Galactic thick disc \citep{Hawkins2015}.
The majority of such halo stars with $-0.7 < \feoh < -0.2$ are argued to be the remnant of the thick disc component as a result of the last major merger event \citep{Belokurov2020}.
These studies imply that the stars with $\feoh > -0.2$ have different origins and have experienced very efficient metal enrichment process like the scenario proposed here.
However, it is unlikely that such an event produce even more metal-rich stars than the thick disc stars.

\begin{figure}
 \begin{center}
  \includegraphics[width=80mm]{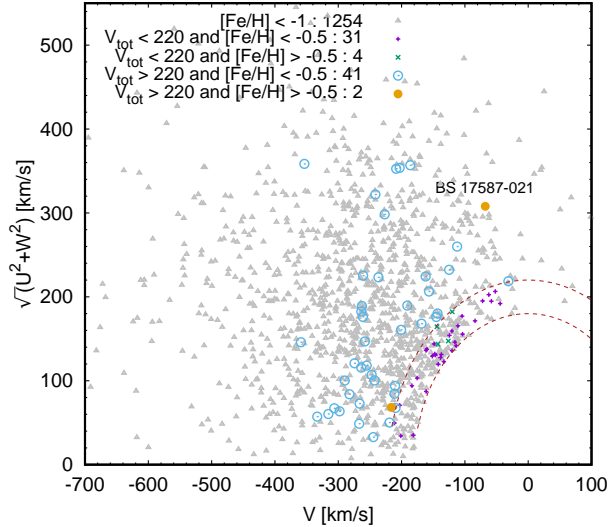}
 \end{center}
 \caption{The Toomre diagram for the stars with space velocities available in the SAGA database. The Galactic coordinate is based on J2000 and is centered on the Sun. The space velocity, $U$, $V$, and $W$ is the heliocentric velocity component with respect to the Galactic centre, the azimuthal angle along the Galactic disc, and the Galactic north pole, respectively. The dashed lines correspond to the total velocity of 180 and 220 km/s, based on the boundary criterion of halo component suggested by the previous studies \citep{Nissen2010,Bonaca2017}.}\label{fig:kin}
\end{figure}

To identify the observational counterparts of the proposed scenario, we performed a cross match of Galactic stars between the SAGA database and Gaia DR2 \citep{Matsuno2019}.
Figure~\ref{fig:kin} shows the sample stars on the Toomre diagram that have the space velocities larger than 180 km/s.
The velocity data are calculated from the proper motion data taken from Gaia DR2 \citep{GaiaCollaboration2018}.
The boundary between the disc and halo stars are defined by the heliocentric velocity, $V_{\rm total} = \sqrt{U^{2} + V^{2} + W^{2}} = 180$ or 220 km/s as shown by the dashed lines in the figure, following the prescription in the previous studies \citep{Nissen2010,Bonaca2017}.
The positions of the stars are calculated based on the right ascension and declination and the distance from parallax in Gaia DR2 where we imposed the criterion for the data accuracy by {\it parallax\_over\_error} $> 5$.
The total number of data with $V_{\rm total} > 180$ km/s is 1332.
Among the stars with $\feoh \geq -1$, 35 stars have $V_{\rm total} < 220$ km/s and 43 stars have $V_{\rm total} > 220$ km/s.
The six stars with $\feoh > -0.5$ are listed in table~\ref{tab:kin}.
Only one star, BS 17587-021 has super-solar metallicity among these metal-rich stars while the others have $\feoh \lesssim -0.4$.
All of the 78 metal-rich stars with $V_{\rm total} > 180$ km/s show retrograde motion.

\begin{table*}
  \caption{List of metal-rich halo stars with $\feoh > -0.5$. The column represents the star name, metallicity, lithium abundance, the space velocities, the total heliocentric velocities, and the chemical compositions of magnesium, silicon, and calcium.}\label{tab:kin}
  \begin{center}
    \begin{tabular}{*{10}{l}}
    \hline
Star Name & \feoh & \ali & $U$ (km/s) & $V$ (km/s) & $W$ (km/s) & $V_{\rm total}$ [km/s] & [Mg/Fe] & [Si/Fe] & [Ca/Fe] \\
 &  &  &  [km/s] & [km/s] & [km/s] & [km/s] &  &  & \\
BS17587-021 & 0.93 & n/a & -207 & -68 & -228 & 315 &  &  & 0.08 \\
HD 107773 & -0.39 & n/a & -142 & -126 & 38 & 194 & 0.29 & 0.25 & 0.21 \\
HIP 71019 & -0.4 & n/a & 120 & -120 & -137 & 218 & 0.25 & 0.23 & 0.18 \\
HD 159001 & -0.44 & n/a & 146 & -144 & -137 & 218 & 0.24 & 0.41 & 0.21 \\
HD 237822 & -0.45 & n/a & 73 & -143 & 124 & 203 & 0.33 & 0.3 & 0.3 \\
HD 180928 & -0.48 & n/a & 64 & -216 & -24 & 227 & 0.38 & 0.27 & 0.15 \\
     \hline
     \end{tabular}
  \end{center}
\end{table*}

The massive binary scenario is a supplementary scenario to the existing three major scenarios for the origin of the known EMP stars, i.e., the scenario is consistent with the current framework.
The three scenarios, (1) the binary mass transfer scenario, (2), the mixing and fallback in the first supernovae scenario, and (3) the fast-rotating massive star scenario, can be tested by surface lithium abundances, especially in scenarios (2) and (3).
The degree of the contribution of a previous generation star to the abundance patterns of an observed star is measured by the dilution factor \citep{Meynet2010}, which is defined by the mass ratio of the ISM to the ejecta in the previous generation star.
The massive binary scenario can reproduce a variation in the dilution factor, by estimating the ratio of the envelope mass to the accreted mass.
Therefore, the models of peculiar supernovae and possibly rotating massive stars may fit with the proposed scenario to explain all the abundance patterns, including lithium.
In particular, the effect of winds from rotating massive stars is worth investigating because the winds are much slower than supernova ejecta and hence the companions can accrete more materials from the winds.

A connection with the binary mass transfer scenario is intriguing. One of the most iron-poor stars HE 0107-5240 belongs to a binary system whose binary period is more than 10 years \citep{Arentsen2019}, which is consistent with the theoretical prediction of the binary mass transfer scenario \citep{Suda2004}.
If long-period binaries are responsible for CEMP stars at low metallicity, a variety of CEMP stars and other metal-poor stars are expected, depending on binary parameters.
Also, we may imagine a more complex scenario for the origin of CEMP stars.
If we try to apply the massive binary scenario to HE 0107-5240, we will need a triple system to complete the abundance pattern.
The accretion of iron and lithium, initiated by the collision of a supernova explosion onto the observed stars is followed by the accretion of carbon and possibly {\it s}-process elements keeping the abundance of lithium and iron-group elements on the surface.
Considering the third star to reproduce all the abundances, we need a fine-tuning of model parameters, while the formation of a triple system is not so rare as seen from the observations of nearby main-sequence stars \citep{Moe2019}.

The formation of binary systems in the early universe will modify the current framework of the history of star formation and chemical evolution in the early universe.
So far, only the first supernovae are thought to have contributed to the early chemical enrichment.
There are many arguments for the formation of second-generation stars associated with the simulations of the first supernovae in the early universe.
The major concerns are the metal mixing of supernova ejecta with the ISM \citep{Whalen2008,Smith2015,Chiaki2020}, the formation of the low-mass, second-generation stars with dust cooling \citep{Yoon2016,Chiaki2017}, the initial mass function of the first and second-generation stars \citep{Hirano2014}, and the chemical yields from the first supernovae \citep{Heger2010,Tominaga2007}.
All of these topics will influence the proposed scenario in favor or disfavor of it.

Our proposed scenario also influences the argument of the initial mass function of the first-generation stars.
Our scenario is consistent with most of the recent simulations on the formation of first-generation stars in which massive stars are dominated.
It will be more important to consider the formation channel of low-mass companions around massive stars.
It is in line with the argument that some low-mass first-generation stars are formed in the disc of the central massive stars \citep{Susa2014}.
The formation of intermediate and massive stars in the early universe is also favored to explain the large fraction of CEMP stars among EMP stars under the binary mass transfer scenario \citep{Komiya2007}.
It is proposed that massive stars dominate the EMP population as a byproduct of the need for many intermediate-mass stars to pollute low-mass companions.

The proposed scenario may provide an important clue to the early chemical enrichment of neutron-capture elements, like barium in EMP stars.
We argue that the first massive stars are the first polluters with metals of low-mass companions, which may include the enrichment of neutron-capture elements.
Observationally, the majority of observed EMP stars exhibit detectable barium absorption lines.
This means that EMP stars will possess the evidence of neutron-capture processes such as the {\it r}-process and the {\it s}-process at any metallicity ranges.
Recent observations and theories prefer neutron-star mergers as the site of the {\it r}-process \citep{Abbott2017} in reproducing the early chemical enrichment of neutron-capture elements in dwarf galaxies in the Local Group (see, e.g. \cite{Ji2016b,Roederer2016} for observations and \cite{Hirai2015,Safarzadeh2017,Tarumi2020} for simulations).
However, it is not sufficient to reproduce the even lower abundances of neutron-capture elements at the lowest metallicity range \citep{Tsujimoto2017}.
Because the {\it s}-process is not expected in supernovae, we may need the production of {\it r}-process elements in supernovae as argued earlier \citep{Wanajo2013}. 

\section{Conclusions}\label{sec:conclusion}

We have proposed a new scenario to find the survivors of the first stars in the local universe through the simulations of the collision between the ejecta of a first supernova and a low-mass star in a close binary system.
The supernova ejecta do not have significant effect on the stripping of the surface layers of low-mass stars at a distance of $\gtrsim 0.1$ au.
The effect of the accretion of ejecta onto the companion star strongly depends on its binary separation.
If the separation is small enough, namely $\lesssim 0.1$ au, the companion star can be very metal-rich, although it depends on the chemical composition of the yield and how the supernova ejecta mix with the convective envelope of low-mass stars. 

We have argued that survivors of the first stars can be found among metal-rich stars in the Galactic halo.
This is due to the small radii in the first massive stars, which enables them to contaminate low-mass companion stars with supernova ejecta in binary systems.
The existence of these stars in former massive binaries is in line with some theoretical models of the formation of the first stars.
We also investigated the literature dealing with binary stars for OB stars in the Galactic disc to confirm the existence of short-period binaries consisting of massive stars and low-mass stars.
Althogh nearby OB stars are metal-rich and the formation mechanism of binary systems must be different, we find potential candidates of the metal-rich counterparts of the progenitor binaries that match with our proposed scenario.
We also looked for metal-rich halo stars in the sample of known Galactic stars with kinematic information considered taken from the Gaia DR2.
There are several stars with $\feoh > -1$ that have space velocities of typical halo population.

The available data suggest that we may already have a number of the first stars in our Galaxy, whose surfaces are contaminated by the ejecta of the first supernovae.
It is encouraged to explore various possibilities of binary formation in the early universe, first supernovae and/or other source of ejecta from massive stars like a wind from a fast-rotating massive star, dynamical process of the collistion between the ejecta and stars.
The detailed chemical abundances and kinematic information of metal-rich halo stars will provide a better constraint on the proposed models.

\section*{Funding}
This work is supported by a Grant-in-Aid for Scientific Research (KAKENHI) (JP16K05287, JP15HP7004, JP16H02166, JP16H06341, JP19HP8019, JP20HP8012) from the Japan Society for the Promotion of Science.
 
\ack{We thank an anonymous referee for checking the manuscript carefully and giving us useful comments which helped to improve the manuscript.
The authors are grateful to Takashi Yoshida and Naoto Kuriyama for a fruitful discussion on stellar evolution models, to Ataru Tanikawa for his comments on the SPH simulations on colliding supernova ejecta on binary companions, and to Wako Aoki for his useful comments on abundance data and kinematics.
We thank Paul Gaysford for his proofreading of the manuscript.
Numerical computations were carried out on Cray XC50 at Center for Computational Astrophysics, National Astronomical Observatory of Japan and Cray XC30 at Earth-Life Science Institute, Tokyo Institute of Technology.
This work is partly supported by the observations using Nayuta telescope operated by Nishi-Harima Astronomical Observatory, Center for Astronomy, University of Hyogo.
}

\bibliographystyle{apj}
\bibliography{apj-jour,reference,reference_mps_obs,reference_dsph}

\section*{Appendix}
\subsection*{The data source for elemental abundances}
In plotting the data points in figure~\ref{fig:life}, we have excluded some references for which the number of data is small.
On the other hand, we have included papers that report lithium abundance only for one star so that the impression from the figure does not change by reducing the data points.
The data in figure~\ref{fig:life} are taken from the following literature: \citet{Fulbright2000,Gratton2000,Ryan2001b,Cowan2002,Sneden2003,Christlieb2004b,Boesgaard2005,Charbonnel2005,Spite2005,Takeda2005,Asplund2006,GarciaPerez2006a,Sivarani2006,Boesgaard2007,Norris2007,Aoki2008,Frebel2008,Roederer2008,Smiljanic2009,Tan2009,Aoki2010,Melendez2010,Sbordone2010,Caffau2011,Hollek2011,Ruchti2011,Bonifacio2012,Masseron2012,Mishenina2012,Keller2014,Roederer2014a,Roederer2014b,SiqueiraMello2014,Bonifacio2015,Frebel2015a,Hansen2015,Jacobson2015,Li2015,Placco2015,Caffau2016,Matsuno2017a,Matsuno2017b,Bonifacio2018,Li2018,Sakari2018,Starkenburg2018}.

\subsection*{Resolution Studies}

\begin{figure}
 \begin{center}
  \includegraphics[width=80mm]{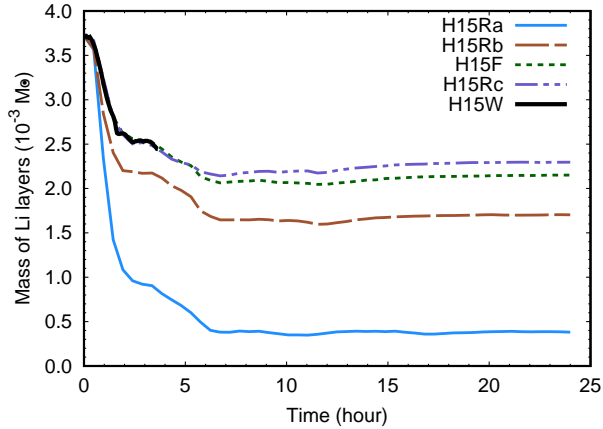}
 \end{center}
 \caption{Time evolution of the mass of lithium-containing layers. Five different mass resolutions, H15F, H15Ra, H15Rb, H15Rc, and H15W were adopted for the ejecta model. The initial separation was fixed to 0.1 au.}\label{fig:stripped}
\end{figure}

Figure~\ref{fig:stripped} shows the time evolution of the mass of lithium-containing layers, which represents the unbound mass by the collision of supernova ejecta, for four models with different mass resolutions in their ejecta.
The three best high-resolution runs (H15W, H15Rc, and H15F) produce very similar results, and hence the H15F model is selected as a fiducial model.
We tried with the highest resolution (H15W), but it took too much time, and we stopped computation at $\sim 4$ hours in the simulation time.
The mass of lithium-containing layers for the H15W model agrees well with the other two high-resolution models, although the comparison can be made only for the early phase of the simulation.
This confirms that anisotropic particle-mass distribution is a good prescription to simulate the collision of ejecta with a target object.
In the lowest resolution case, the unbound mass is larger compared to the other runs, which may come from the coarse sampling of the ejecta.
Therefore, we adopted the run with $N_{\rm p,ej} =$ 4,800,569 and $m_{\rm p,ej}$, i.e., $\pow{1}{-7} \msun$ as the fiducial resolution in this study.

\subsection*{Dependence on SPH scheme}\label{sec:dependence}

Previous studies point out that the conventional formulation of SPH is not suitable to handle the contact discontinuity and is not able to detect some fluid instabilities \citep{Agertz2007,Ritchie2001,Okamoto2003}.
To overcome the difficulties in treating the shock properly, many efforts have been made to improve the prescription and/or to invent a new scheme of SPH \citep{Saitoh2013,Price2008,Read2010,Hosono2013,Hopkins2013,Yamamoto2015}.
We compared the results using different SPH schemes, i.e., the conventional SPH and DISPH, under the same configuration. 

\begin{figure}
 \begin{center}
  \includegraphics[width=80mm]{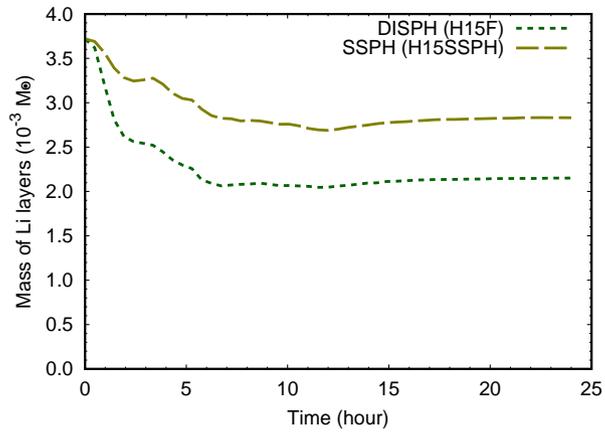}
 \end{center}
 \caption{Same as figure~\ref{fig:stripped}, but for different schemes (H15F and H15SSPH)}\label{fig:dependence}
\end{figure}

Figure~\ref{fig:dependence} compares the conventional (standard) SPH (SSPH) and the DISPH with the same initial condition.
The amount of unbound mass due to the stripping is less in the conventional SPH model than in the DISPH model.
This can be interpreted as a consequence of introducing an unphysical surface tension at the contact surface between the star and the ejecta, which suppresses the stripping of the SPH particles.

\end{document}